# Mobile Health Text Misinformation Identification Using Mobile Data Mining

Wen-Chen Hu
*University of North Dakota, USA*

Sanjaikanth E Vadakkethil Somanathan Pillai
*University of North Dakota, USA*

Abdelrahman Ahmed ElSaid
*University of Puerto Rico at Mayagüez, Puerto Rico*

## ABSTRACT

*More than six million people died of the COVID-19 by April 2022. The heavy casualties have put people on great and urgent alert and people try to find all kinds of information to keep them from being inflected by the coronavirus. This research tries to find out whether the mobile health text information sent to people's devices is correct as smartphones becoming the major information source for people. The proposed method uses various mobile information retrieval and data mining technologies including lexical analysis, stopword elimination, stemming, and decision trees to classify the mobile health text information to one of the following classes: (i) true, (ii) fake, (iii) misinformative, (iv) disinformative, and (v) neutral. Experiment results show the accuracy of the proposed method is above the threshold value 50%, but is not optimal. It is because the problem, mobile text misinformation identification, is intrinsically difficult.*

## INTRODUCTION

In the past, people received their information through various communication channels such as TVs, radios, newspapers, and magazines. However, since the extremely high prevalence of smartphones, human's life style has been changed significantly. There are more than one billion smartphones sold worldwide each year. People rely on smartphones to perform their daily activities like texting, checking messages, watching news, playing games, paying fees, etc. Mobile messages become the major information and news source for many people. Unfortunately, many messages received on the devices may not be correct. Incorrect information and news from the traditional channels are a big headache for societies. The problem is even worse for mobile messages since the messages could come from anywhere, individuals or organizations, instead of few fixed sources. Some of the messages may not be correct and are delivered intentionally or unintentionally. For example, the claim of election fraud with mail-in ballots has many people lose their faith on election justice. This claim is arguable, but at least should not be broadcasted without further proof, and it may be sent intentionally to affect the election results. This kind of mobile misinformation could have huge impacts on societies and needs to be identified and stopped. The consequences of health or medical misinformation are even worse than the ones of generic misinformation because the former may be fatal. For example, the misinformation of eating garlic able to prevent COVID-19 may lead people to lower their vigilance against coronavirus. Figure 1 shows the possible impacts of misinformation during a pandemic.

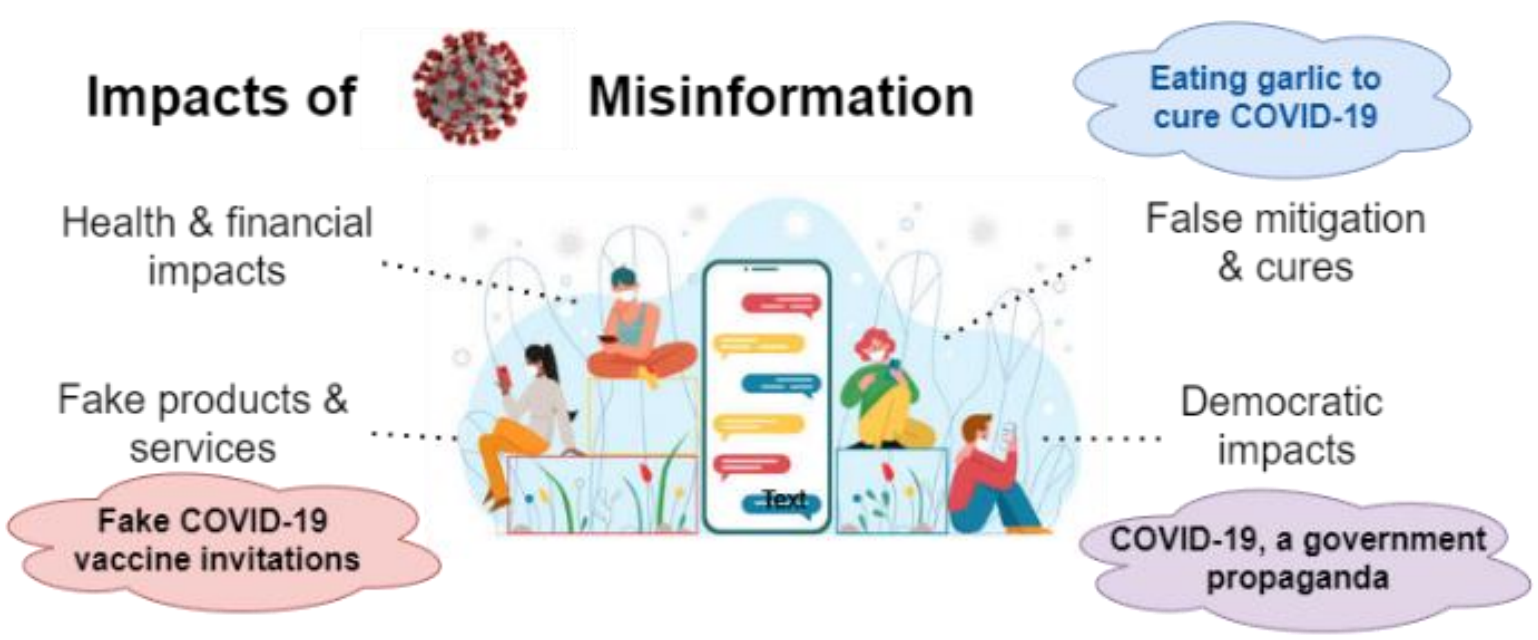

*Figure 1. Possible impacts of misinformation during a pandemic*

Data mining is an effective tool to fight misinformation (Horowitz, 2021). This research tries to propose a method for effectively identifying mobile health misinformation, especially pandemic misinformation, sent to users' devices by using various mobile information retrieval and data mining technologies including lexical analysis, stopword elimination, stemming, and decision trees. The mobile text messages are classified into five classes: true, fake, misinformative, disinformative, and neutral. Users could make better judgement regarding the messages after consulting the classes found by our method. For example, if a text message is labeled as fake, the user may ignore it just by browsing its title without reading the details. Experiment results show the accuracy of the proposed method is above the threshold value 50%, but is not optimal. It is because the problem, health misinformation identification, is intrinsically difficult and the text messages do not provide much information to dig. Further refinements are needed before it is put to actual work.

The rest of this paper is organized as follows. Section 2 presents related works on misinformation identification. The system structure and its components are given in Section 3. Section 4 proposes our major work, a decision tree, for detecting mobile health text misinformation. The experiment results and discussions are given in Section 5, followed by a conclusion including future research directions and references.

## BACKGROUND AND RELATED LITERATURE

This section gives the background information of this research and related research in case readers are interested in finding more relevant publications. Many individuals receive information from their devices in these days, and people tend to trust the information blindly without realizing the data could be wrong. Researchers believe that the world is not only fighting Coronavirus but also dealing with "infodemic." In accordance with e-survey conducted by Gupta, Gasparyan, Misra, Agarwal, Zimba, & Yessirkepov (2020), common sources of COVID-19 information are largely from journals and social media platforms. This survey has 128 respondents from different parts of the world with approximately 22 questions to determine the sources of information and misinformation. The authors conclude that television and social media channels are the primary sources of misinformation in terms of COVID-19. A different article provides a mixed-method analysis of COVID-19 misinformation (Brennen, Simon, & Nielsen, 2021). The authors provide both text and visual elements of misinformation that are already rated false or misleading in the dataset. They successfully represent aspects and functions of misinformation by uncovering a small number of manipulated visuals using artificial intelligence-based techniques. Related research about coronavirus misinformation identification can be found from the articles (Mian & Khan, 2020; Fleming, 2020; Ball & Maxmen, 2020; Zhou & Zafarani 2020; Khan, Michalas, & Akhunzada, 2021; Savage, 2021).

Misinformation identification and management is critical and popular in these days because information could be created and sent by everyone, not just news agencies, and some may distribute misinformation unintentionally or on purpose. Sitaula, Mohan, Grygiel, Zhou, & Zafarani (2020) detect fake news by assessing its credibility. By analyzing public fake news data, they show that information on news sources (and authors) can be a strong indicator of credibility. Their findings suggest that an author's history of association with fake news, and the number of authors of a news article, can play a significant role in detecting fake news. Their approach can help improve traditional fake news detection methods, wherein content features are often used to detect fake news.

Reddy, Raj, Gala, & Basava (2020) discusses approaches to detection of fake news using only the features of the text of the news, without using any other related metadata. They observe that a combination of stylometric features and text-based word vector representations through ensemble methods can predict fake news with an accuracy of up to 95.49%. Efforts to combat the effects of fake news focus too often exclusively on the factual correctness of the information provided. To counter factually incorrect—or incomplete, or biased—news, a whole industry of fact-checkers has developed. While the truth of information that forms the basis of a news article is clearly of crucial importance, there is another, often overlooked, aspect to fake news. Successfully recognizing fake news depends not only on understanding whether factual statements are true, but also on interpreting and critically assessing the reasoning and arguments provided in support of conclusions. It is, after all, very possible to produce fake news by starting from true factual statements and drawing false conclusions by applying skewed, biased, or otherwise defective reasoning. Visser, Lawrence, & Reed (2020) therefore argue that fact-checking should be supplemented with reason-checking: evaluating whether the complete argumentative reasoning is acceptable, relevant, and sufficient.

Dealing with complex and controversial topics like the spread of misinformation is a salient aspect of our lives. Sethi, Rangaraju, & Shurts (2019) present initial work towards developing a recommendation system that uses crowd-sourced social argumentation with pedagogical agents to help combat misinformation. We model users' emotional associations on such topics and inform the pedagogical agents using a recommendation system based on both the users' emotional profiles and the semantic content from the argumentation graph. This approach can be utilized in either formal or informal learning settings, using threaded discussions or social networking virtual communities. Recent developments in neural language models (LMs) have raised concerns about their potential misuse for automatically spreading misinformation. In light of these concerns, several studies have proposed to detect machine-generated fake news by capturing their stylistic differences from human-written text. These approaches, broadly termed stylometry, have found success in source attribution and misinformation detection in human-written texts. Schuster, Schuster, Shah, & Barzilay (2020) show that stylometry is limited against machine-generated misinformation. Whereas humans speak differently when trying to deceive, LMs generate stylistically consistent text, regardless of underlying motive. Thus, though stylometry can successfully prevent impersonation by identifying text provenance, it fails to distinguish legitimate LM applications from those that introduce false information. Their findings highlight the need for non-stylometry approaches in detecting machine-generated misinformation, and open up the discussion on the desired evaluation benchmarks.

Misinformation identification is not an easy topic to tackle because the classification is usually subjective and the short messages do not provide much information themselves. In addition, a variety of misinformation exists which requires a great amount of data to be stored. Research of misinformation identification has been studied extensively, and this research does not intend to solve all the misinformation problems at once. Instead, it focuses on two themes: short text messages and health-related misinformation, especially the COVID-19. Interested readers can refer to other misinformation identification methods from the articles (Collins, Hoang, Nguyen, & Hwang, 2021; Guo, Ding, Yao,

Liang, & Yu, 2020; Reis, Correia, Murai, Veloso, & Benevenuto, 2019; Roozenbeek & Linden, 2019; Zhou & Zafarani, 2020; Aldwairi & Alwahedi, 2018).

## THE PROPOSED SYSTEM

This research tries to identify mobile health text misinformation sent to users' devices by using various information retrieval and data mining technologies. The whole structure of the system and details of its components except the major method, decision trees, are described in this section.

### A Workflow of the Proposed System

Construction of the proposed system includes two steps: training and testing. The training phase builds a decision tree based on the sample pandemic mobile text messages. After the decision tree is built, a forthcoming message can be checked to see whether it is misinformation during the testing phase, and an appropriate action, like showing a warning, will be taken if the message is deemed misinformation. In order to better explain our system, Figure 2 shows a workflow of the proposed system, which includes the following five methods/components:

- *Lexical analysis*, which is the process of converting the input stream of characters of a message into a stream of words. For example, it breaks the string "The COVID-19 vaccines were not properly tested or developed." into nine words "The," "COVID-19," "vaccines," "were," "not," "properly," "tested," "or," and "developed."
- *Stopword removal*, which is to remove the trivial words, like a, and, and so, from the message. For example, this phase drops the the stopwords "The," "were," "not," and "or."
- *Stemming*, which is the process for reducing inflected words to their stem, like the three terms: went, gone, and going, converted to the term: go. For example, "vaccines" becomes "vaccine," "tested" becomes "test," and "developed" becomes "develop."
- *Database*, which saves critical keywords and phrases constantly for updating the decision tree later.
- *Building a decision tree*, which is used to classify the forthcoming mobile messages into five classes, and the details of this construction and database will be discussed in the next section.

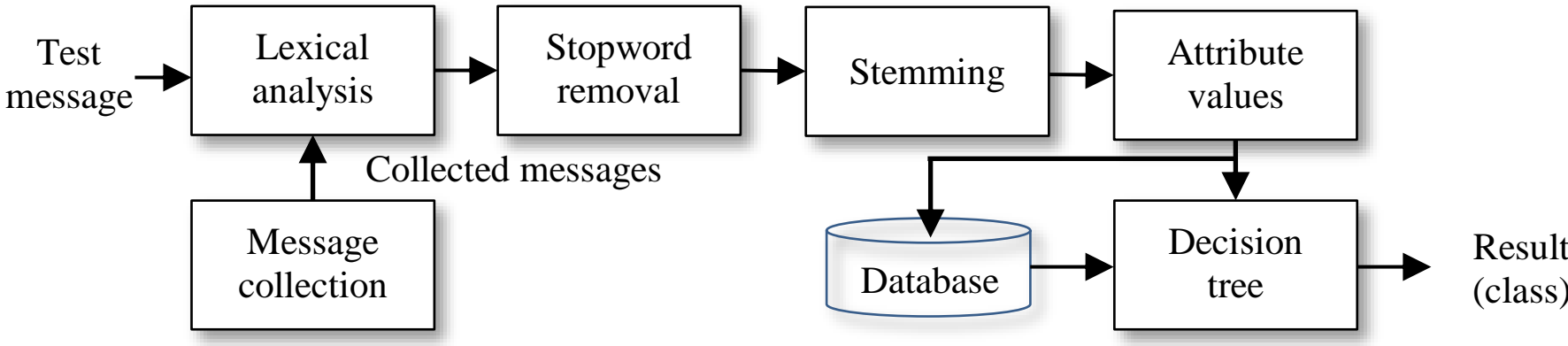

*Figure 2. A workflow of the proposed system for identifying mobile health misinformation*

### The Mobile Data/Text Mining Technologies Used

People are flooded by a plethora of mobile data every day. Some of the data is relevant or important, but some of it is irrelevant or even harmful. Similarly, not all the collected data is relevant while training the system. So, all the hyperlinks, non-alphanumeric character and English stopwords such as has, is, are, etc., should be removed from the dataset before analyzing the dataset. The primary text processing steps are given next. Though the methods applied are classical, they are essential for this research. Interested readers may refer to the article by Stavrianou, Andritsos, & Nicoloyannis (2007, September).

#### *Lexical Analysis*

Lexical analysis is the process of converting an input stream of characters into a stream of words or tokens, which are groups of characters with collective significance. It is the first stage of automatic indexing which is the process of algorithmically examining information items to generate lists of index terms. The lexical analysis phase produces candidate index terms that may be further processed, and eventually added to indexes. It also helps split the longer sentences into smaller chunks of the dataset to perform algorithms with better accuracy.

### *Removal of Stopwords*

English stopwords such as is, has, an, the, etc. do not signify any importance as index terms when analyzing the dataset for information. It is crucial to remove the stopwords from the dataset as they do not help us find the true meaning of a sentence and can be removed without any negative consequences. Also, eliminating such words from consideration early in automatic indexing speeds processing, saves huge amounts of space in indexes. It has been recognized since the earliest days of information retrieval that many of the most frequently occurring words in English (like "the," "of," "and," "to," etc.) are worthless as index terms. A search using one of these terms is likely to retrieve almost every item in a database regardless of its relevance, so their discrimination value is low. Furthermore, these words make up a large fraction of the text of most documents: the ten most frequently occurring words in English typically account for 20 to 30 percent of the tokens in a document. Eliminating such words from consideration early in automatic indexing speeds processing, saves huge amounts of space in indexes, and does not damage retrieval effectiveness.

### *Stemming*

It is a technique for improving retrieval effectiveness and reducing the size of indexing files is to provide searchers with ways of finding morphological variants of search terms. If, for example, a searcher enters the term stemming as part of a query, it is likely that he or she will also be interested in such variants as stemmed and stem. Since a single stem typically corresponds to several full terms, by storing stems instead of terms, compression factors of over 50 percent can be achieved. The stem need not be identical to the morphological root of the word; it is usually sufficient that related words map to the same stem, even if this stem is not in itself a valid root. It is a method for casting words into their original form which aims to the removal of inflectional endings from words. It performs morphological analysis on the words by returning the words into its dictionary meaning. For example, the stemming converts caring into care, troubled into trouble, geese into goose, etc.

### *Databases*

In addition, some attribute values will be saved in a database for revising the decision tree constantly and later. For example, if a phrase like "Johnson & Johnson" has been appearing in true messages many times, it may be added to the database for rebuilding the decision tree later. Database sample values are shown in Table 1, which consists of two tables. One saves the properties of the attributes, and another table stores the attribute values and their counters.

*Table 1. Database tables to save critical message attributes for revision: (a) attribute properties and (b) attribute values and their counters*

| ID | attribute | value |
|---|---|---|
| 1 | source | text |
| 2 | negative | keywords |
| 3 | positive | keywords |
| 4 | time | date and time |
| 5 | spamming | Boolean |
| 6 | multiple | Boolean |
| 7 | spamming | Boolean |
| 8 | advertisement | Boolean |

(a)

| ID | attribute value | counter |
|---|---|---|
| 1 | CDC | 28 |
| 1 | https://www.cnn.com/ | 12 |
| 1 | White House | 14 |
| 2 | credit card | 5 |
| 2 | sign in | 6 |
| 3 | vaccine | 36 |
| 3 | Fauci | 16 |

(b)

## THE DECISION TREE USED BY THE PROPOSED METHOD

Text classification can be achieved by using various methods including machine learning algorithms. Decision trees are a supervised learning classification technique that can be used to predict a pattern or classify a data set into classes. It uses the branch method to determine the final results by breaking down the data set into smaller and smaller subsets of the data.

### Decision Trees

A decision trees is a hierarchical tree structure that can be used to classify data based on a series of questions about the attribute values of the data (Jenhani, Amor, & Elouedi, 2008). The attributes of the data can contain any values such as binary, ordinal, categorical, nominal, etc. Figure 3 shows a generic decision tree which is a supervised learning consisting of two steps, training and testing. Various COVID-19 text messages are collected from the Internet and are used to build a decision tree. To classify the data, the testing starts from the root node. Based on the attribute values of the data, child node is selected until reaching a leaf which gives the class the data belongs to.

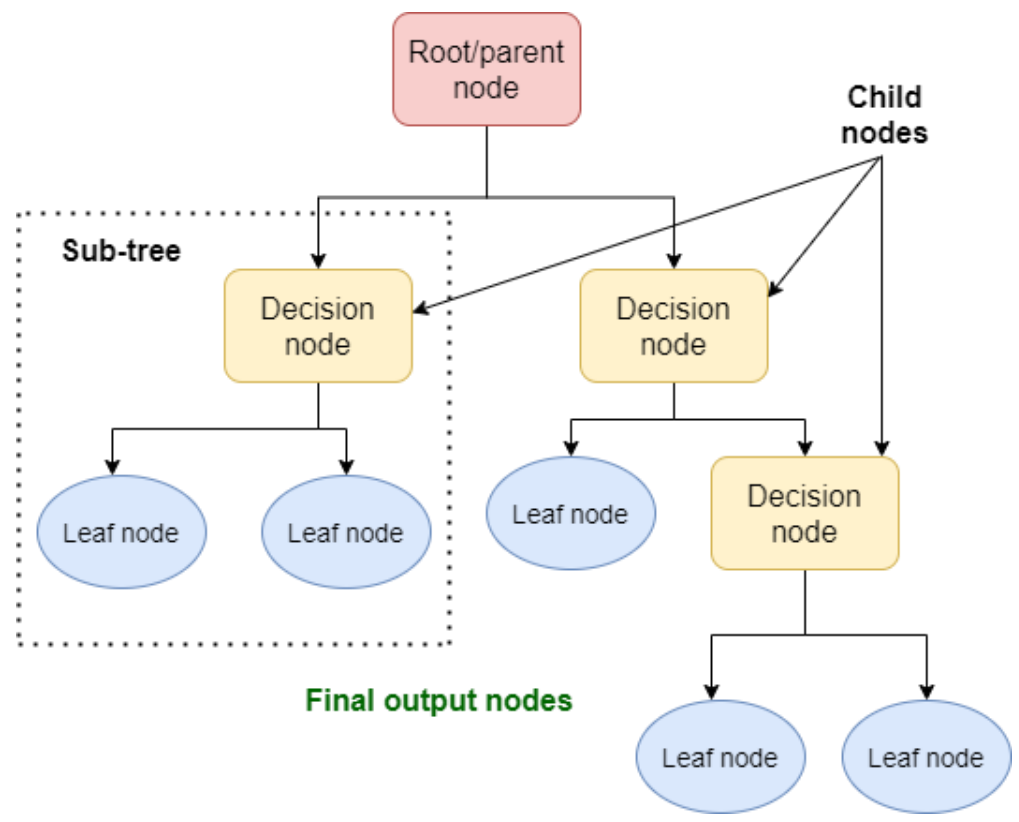

*Figure 3. A generic decision tree*

### Algorithm of Decision Tree Building

The method of decision trees is classical and traditional, but it is effective and is applied to identify the health misinformation in this research. The algorithm of decision tree construction is given in this section (Buhrman & de Wolf, 2002; Bombara, Vasile, Penedo, Yasuoka, & Belta, 2016). To help readers better understand our method, two representations of the proposed method are given in this section. Figure 4 gives a flowchart of the decision tree construction, and the associated algorithm is given in Figure 5. To decide how to split the set, the impurity is used to measure the homogeneity of a data set. The impurity of the subset of an attribute used in this research is based on the following entropy:

$$Impurity = \sum_{i=1}^{h} -p_i \times log_2 p_i$$

where $p_i$ is the value of probability of a class $i$ and $h$ is the number of classes. The measure to compare the difference of impurity degrees is called information gain (IG), and is found by using the following formula for an attribute:

$$Information\ gain = Entropy\ of\ data\ set\ S - \sum_{j=1}^{m} \frac{k}{n} \times Entropy\ of\ each\ value\ k\ of\ subset\ S_j$$

where $S_j$ is the subset with a specific attribute value, $m$ is the number of different attribute values, $n$ is the number of attribute values of the set $S$, and $k$ is the number of attribute values of the subset $S_j$. The attribute with the maximum information gain is used to split the data set $S$ into several subsets $S_k$. A decision tree is found by continuing to split the data set until no subset has different classes.

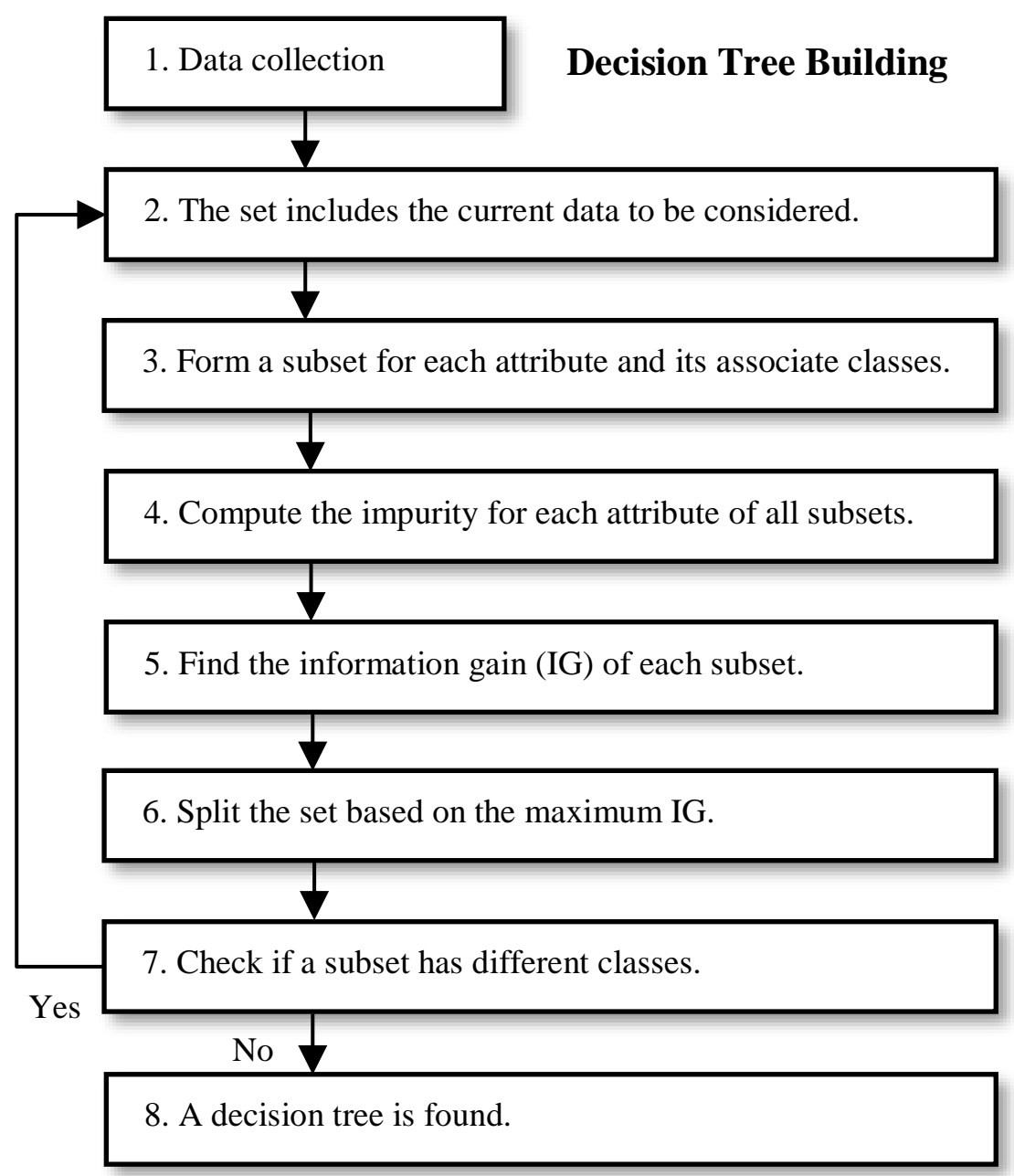

*Figure 4. A flowchart of the decision tree construction*

***BUILD-DECISIONTREE**( d )*

***Input:** Attributes & classes from sample messages, d*
***Output:** A decision tree*

1. ***loop***
2. *maxGain ← 0*
3. *maxA ← null*
4. *s ← d*
5. *i[attribute] ← FIND-IMPURITY( s, s[attribute] )*

```
6.     for each  e  in  i[attribute]  do
7.         gain ← FIND-INFORMATIONGAIN( s, e )
8.         if  gain > maxGain  then
9.             maxGain ← gain
10.            maxA ← e
11.        end if
12.    end for
13.    d ← SPLIT-SET( s, maxA )
14. until all subsets processed
```

*Figure 5. Algorithm of the decision tree construction*

Each set of data will create a unique decision tree. Figure 6 shows an example of a decision tree may be created by using our method to help readers understand the above flowchart and algorithm.

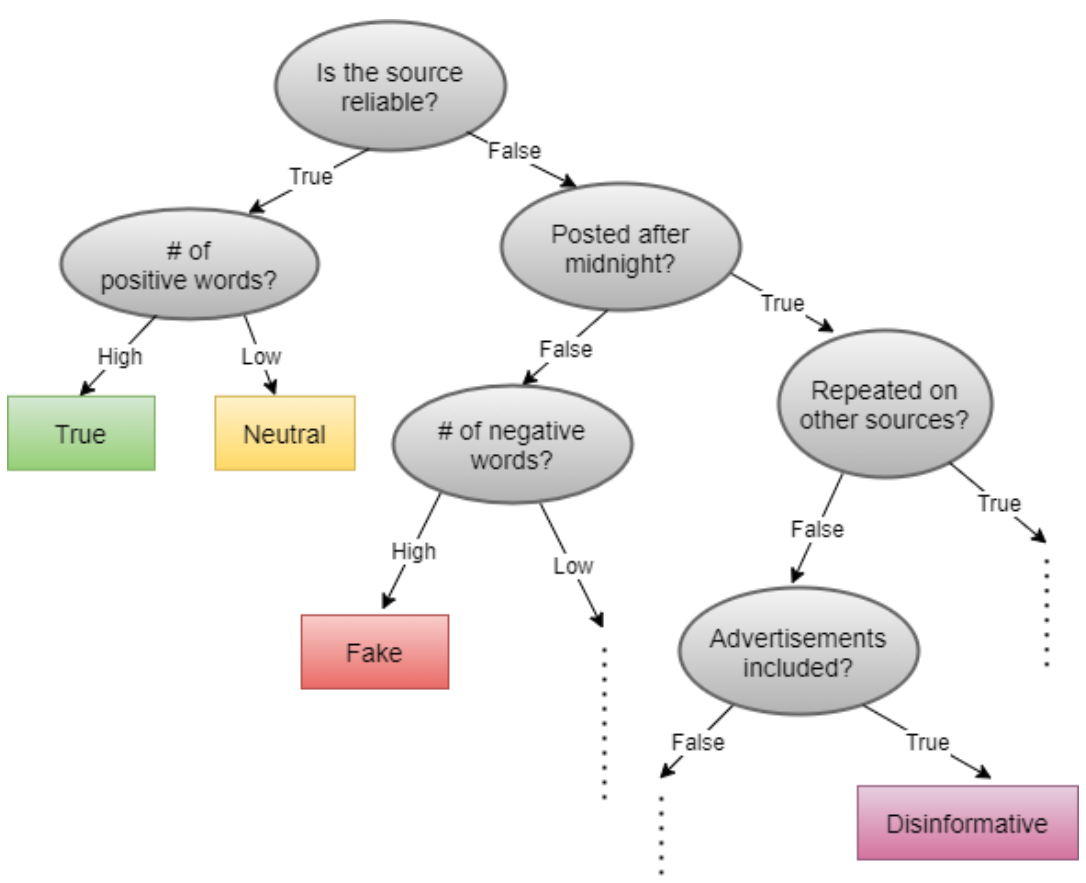

*Figure 6. A sample of a decision tree built by using our method*

## The Attributes and Final Classes

The data can be collected from various websites containing posts, news, and articles consisting of real and fake data. In order to use the data for analysis and prediction or detection of misinformation using machine learning algorithms, it needs to be preprocessed. Various attributes from the dataset can be used for better accuracy. The dataset can consist of the source, date and time of the information to ensure reliability. For example, the system can rank each source like CDC and WHO with the highest rating and individuals with no verified handles, friends, and businesses with low-ranking of reliability. Also, the attributes like repeated words, positive and negative keywords, news with lots of advertisements can help detect misinformation. The decision tree created and tested is based on the attribute values of the mobile messages including:

- *Source*, which is the source sending the message. Each source has a ranking in our database. For example, the CDC has the highest ranking, and friends & businesses have low rankings.
- *Negative keywords or phrases*, which have a negative impact if including keywords like credit card, money, fund, etc.
- *Positive keywords or phrases*, which have a positive impact if keywords like CDC, hospitals, government, etc. are included

In addition, the above attribute values can be updated automatically from time to time based on the test data. For example, if a source keeps sending reliable messages, it could be added to the database with a good ranking. On the other hand, the following attribute values are not able to be updated:

- *Date and time*, which have a negative impact if the message is sent after midnight, for example,
- *Multiple sources*, which have a positive impact if it is sent by multiple reliable sources,
- *Spamming*, which has a negative impact if multiple copies of the message is sent, and
- *Advertisements*, which have a negative impact if advertisements are included.

A mobile health text message can be classified into the following five classes:

- *True*, which is true information and is without a doubt. For example, it is true that a vaccine to prevent COVID-19 is available because COVID-19 vaccines have been authorized by the U.S. Food and Drug Administration (FDA) and vaccine programs have begun across the country.
- *Fake*, which could be either misinformation or disinformation. For example, it is an obviously fake news that the COVID-19 vaccines contain microchips for government tracking because the current technology has not been this advanced yet.
- *Misinformative*, which is false or out-of-context information that is intentionally or unintentionally presented as fact to deceive. For example, it is misinformative that the COVID-19 vaccines are mandatory because they are strongly recommended, but not mandatory.
- *Disinformative*, which is a type of misinformation that is intentionally delivered the false or misleading information to deceive or mislead readers. For example, it is disinformation that the COVID-19 vaccines were not properly tested or developed since they were tested legitimately.
- *Neutral*, which cannot be decided by the proposed method. For example, our method is not able to decide whether Coronavirus is from labs since even the societies have controversy about this in these days, let alone software.

The differences between misinformation and disinformation are not distinct. This research treats the former as a mistake. If the information is intentionally to deceive, it is classified as disinformation. Otherwise, it is misinformation. That is misinformation actively demonstrates the information that is communicated to mislead, whereas disinformation can be recognized as malicious tricks and computational publicity.

## EXPERIMENT RESULTS

This research is related to mobile computing, so the proposed method is tested via an app. This section shows the experiment results of our method and gives discussions about the results.

### Experiments

Figure 7 shows the experiment setup for testing our system. A decision tree actually could be located at either client or server, but its construction should be done at the server for convenience. The database is for saving and updating the keywords for revising the decision tree later. Also, the following Android system is used for testing:

- Android Studio 4.13,
- Android 11.0 (R) Platform (API 30), and
- Android Virtual Device (AVD) (API 30 and 3.2” GVGA).

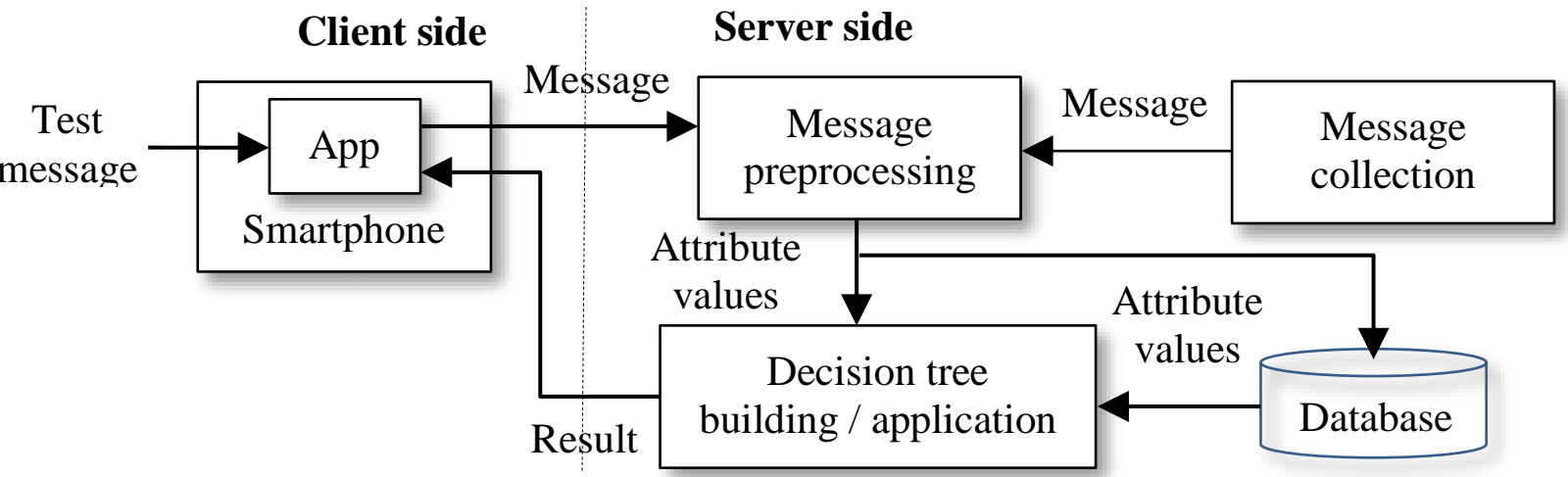

*Figure 7. Another view of the proposed system for experiments*

Figure 8.a shows the proposed app displayed on the Android launcher window. After the user clicking on the app icon, a list of text messages is shown in Figure 8.b, and the user can pick one to read. Figures 8.c and 8.d show the results after checking two messages.

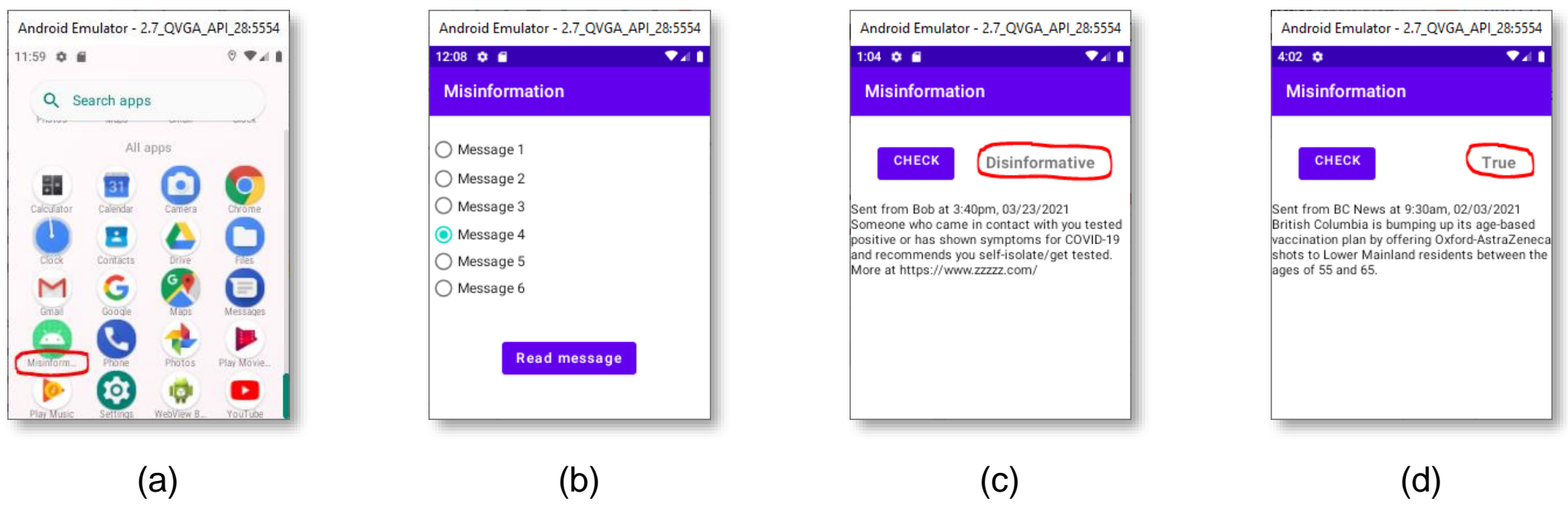

*Figure 8. (a) the app on an Android launcher, (b) selecting which message to check, (c) checking the message and showing the class, and (d) checking another message*

## Evaluations and Discussions

Table 2 shows the evaluation data from testing 100 mobile pandemic messages randomly collected from various sources such as the Internet and online short message archives (COSMOS, n.d.). Typical messages are given as follows:

- *True*: "As of March 28, 2021, 30+ million US cases of COVID-19 were reported to CDC. Cases are rising again. The 7-day average of new daily cases is over 60,000, over a 10% increase from the previous week," from CDC on 03/28/2021.
- *Fake*: "A retired anesthesiologist who rails against mask wearing to protect against COVID-19 purports to demonstrate their ineffectiveness by blowing out the vapor from an electronic cigarette," from Facebook on 02/08/2021.
- *Misinformative*: "80% of people taking the #Maderna vaccine had significant side-effectives via #BillGates," from Facebook on 03/25/2021.
- *Disinformative*: "A nurse pretending to vaccinate actor Anthony Hopkins before squirting the fluid from the syringe on the floor," from Instagram on 03/10/2021.
- *Neutral*: "The vaccine was 100% effective against severe disease as defined by the CDC, and 95.3% effective against severe COVID-19 as defined by the FDA," from Pfizer and BioNTech on 03/31/2021.

*Table 2. Evaluation data from testing 100 mobile text messages*

| Message Type | # of Messages | # of Correct Detection | Accuracy |
|---|---|---|---|
| True | 40 | 25 | 63% |
| Fake | 20 | 7 | 35% |
| Misinformative | 10 | 5 | 50% |
| Disinformative | 10 | 5 | 50% |
| Neutral | 20 | 15 | 75% |
| Overall | 100 | 57 | 57% |

Identifying misinformation is intrinsically difficult. People are not able to tell whether the information is correct easily, let alone computers. The accuracy of the proposed method is above the threshold value, 50%, but is not impressive. However, this research is not without some merits. For example, it provides the following advantages:

- This is unique research about the mobile health text misinformation identification, which uses various mobile information retrieval and data mining technologies to enhance the mobile research and pandemic prevention.
- Though the tests show the results are not optimal, it is much better than without any help. For example, users can make a better judgement of whether the message is misinformation after consulting the recommendation from our method.
- The low accuracy is mainly due to the lack of information provided by the text messages. To fix the problem, mining the text data, like cross referencing, may be needed other than using the apparent data.
- This research is able to adapt and improve automatically while tests are conducted by updating the data like positive and negative keywords saved in the database. Of course, the decision tree needs to be rebuilt based on the updated data.

In addition, the false alarms may reduce the effectiveness of this research. The future research needs to address this problem too.

## CONCLUSION

Smartphones are indispensable devices for people in these days, and tens or even hundreds of text messages are sent to each device every day. All kinds of information can be found from the delivered messages such as news, greetings from family members or friends, advertisements, promotions, weather reports, etc. People are overwhelmed by the sheer amount of information and they spend much time trying to find a way to sort out the messages. Even worse is some messages give false or fake information and mislead the readers consequently. The problem becomes even more serious especially during the pandemic. This research tries to automatically classify the mobile health text messages into one of the five classes (true, fake, misinformative, disinformative, and neutral) by using various mobile information retrieval and data mining technologies, which include lexical analysis, stopword elimination, stemming, and decision trees. The major method used in this research is a revised decision tree, which may be effective, but may not be innovative enough. Experiment results show the research needs more refinements before it is put into effective use.

To improve the accuracy of the mobile health text misinformation identification, the authors are also considering using other methods such as artificial neural networks (ANN), statistical means, and NLP (natural language processing) to solve this problem, and see whether better results could be found. The ANN is considered because this problem has no definite answers. For example, a message may be considered true for some people, but others may think it is disinformative, especially if it is related to politics, and ANN is competent for this kind of ambivalence (Sinha, Sakshi, & Sharma, 2021; Ahmed, Ali, Hussain, Baseer, & Ahmed, 2021; Isha Priyavamtha, Vishnu Vardhan Reddy, Devisri, & Manek. 2021; Kula, Choraś, Kozik, Ksieniewicz, & Woźniak, 2020). The statistical means includes the methods of Bayesian classifiers and hidden Markov models. It is less innovative, but may be more effective. On the other end, this problem, mobile health text misinformation identification, could be classified as one of the NLP problems (Islam, Liu, Wang, & Xu, 2020). One of the NLP methods is sentence similarity measurement, and various measurements will be adapted to our problem and hope the better results will come up. Besides, there has been a rising interest in proactive intervention strategies to counter the spread of misinformation and its impact on society (Sharma, Qian, Jiang, Ruchansky, Zhang, & Liu, 2019). Methods to mitigate the ill effects caused by misinformation will be investigated too.

## REFERENCES

Ahmed, B., Ali, G., Hussain, A., Baseer, A., & Ahmed, J. (2021, April). Analysis of text feature extractors using deep learning on fake news. *Engineering, Technology, & Applied Science Research*, 11(2), 7001–7005.

Aldwairi, M. & Alwahedi, A. (2018, November 5). Detecting fake news in social media networks. *Procedia Computer Science*, 141, 215-222.

Ball, P. & Maxmen, A. (2020). The epic battle against coronavirus misinformation and conspiracy theories. *Nature*, 581(7809):371+.

Bombara, G., Vasile, C.I., Penedo, F., Yasuoka, H., & Belta, C. (2016, April). A decision tree approach to data classification using signal temporal logic. In *Proceedings of the 19th International Conference on Hybrid Systems: Computation and Control*, 1-10.

Brennen, J.S., Simon, F.M., & Nielsen, R.K. (2021). Beyond (mis) representation: visuals in COVID-19 misinformation. *International Journal of Press/Politics*, 26(1), 277-299.

Buhrman, H. & de Wolf, R. (2002). Complexity measures and decision tree complexity: a survey. *Theoretical Computer Science*, 288, 21-43.

Collins, B., Hoang, D.T., Nguyen, N.T., & Hwang, D. (2021). Trends in combating fake news on social media – a survey, *Journal of Information and Telecommunication*, 5:2, 247-266

COSMOS (Collaboratorium for Social Media and Online Behavioral Studies). (n.d.). *Misinformation*. Retrieved from https://cosmos.ualr.edu/misinformation

Fleming, N. (2020, June 17). Coronavirus misinformation, and how scientists can help to fight it. *Nature*, 583(7814):155-156.

Guo, B, Ding, Y, Yao, L., Liang, Y., & Yu, Z. (2020, September 4). The future of false information detection on social media: new perspectives and trends. *ACM Computing Surveys*, 53(4):68, 1-36.

Gupta, L., Gasparyan, A.Y., Misra, D.P., Agarwal, V., Zimba, O., & Yessirkepov, M. (2020, July 13). Information and misinformation on COVID-19: a cross-sectional survey study. *Journal of Korean Medical Science*, 35(27): e257.

Horowitz, Brian T. (2021, March 15). Can AI stop people from believing fake news? *IEEE Spectrum*.

Isha Priyavamtha, U.J., Vishnu Vardhan Reddy, G., Devisri, P., & Manek. A.S. (2021). Fake news detection using artificial neural network algorithm. In K.R. Venugopal, P.D. Shenoy, R. Buyya, L.M. Patnaik, & S.S. Iyengar (eds), *Data Science and Computational Intelligence*. ICInPro. Communications in Computer and Information Science, Springer, Cham, vol 1483, 2021.

Islam, M.R., Liu, S., Wang, X., & Xu, G. (2020, September 29). Deep learning for misinformation detection on online social networks: a survey and new perspectives. *Social Network Analysis and Mining*, 10:82.

Jenhani, I., Amor, N., & Elouedi, Z. (2008, August). Decision trees as possibilistic classifiers. *International Journal of Approximate Reasoning*, 48(3), 784-807.

Khan, T., Michalas, A., & Akhunzada, A. (2021, September 15). Fake news outbreak 2021: Can we stop the viral spread? *Journal of Network and Computer Applications*, 190, 103112.

Kula, S., Choraś, M., Kozik, R., Ksieniewicz, P., & Woźniak, M. (2020). Sentiment analysis for fake news detection by means of neural networks. In V.V. Krzhizhanovskaya, et al. (eds), Computational Science – ICCS 2020. ICCS 2020. *Lecture Notes in Computer Science*, Springer, Cham, vol 12140.

Mian, A. & Khan, S. (2020, March). Coronavirus: the spread of misinformation. *BMC Medicine*, 18(89).

Reddy, H., Raj, N., Gala, M., & Basava, A. (2020, February 18). Text-mining-based fake news detection using ensemble methods. *International Journal of Automation and Computing*, 17, 210-221.

Reis, J. C. S., Correia, A., Murai, F., Veloso, A., & Benevenuto, F. (2019, March-April). Supervised learning for fake news detection. *IEEE Intelligent Systems*, 34(2), 76-81.

Roozenbeek, J. & Linden, S. v. d. (2019). The fake news game: actively inoculating against the risk of misinformation. *Journal of Risk Research*, 22(5), 570-580.

Savage, N. (2021, March 12). Fact-finding mission. *Communications of the ACM*, 64(3), 18-19.

Schuster, T., Schuster, R., Shah, D.J., & Barzilay, R. (2020). The limitations of stylometry for detecting machine-generated fake news. *Computational Linguistics*, 46(2), 499-510.

Sethi, R.J., Rangaraju, R., & Shurts, B. (2019, May 30). Fact checking misinformation using recommendations from emotional pedagogical agents. In A. Coy, Y. Hayashi, and M. Chang (eds), *Intelligent Tutoring Systems*, 99-104.

Sinha, H., Sakshi, & Sharma, Y. (2021). Text-convolutional neural networks for fake news detection in Tweets. In V. Bhateja, SL. Peng, S.C. Satapathy, and YD. Zhang (eds), *Evolution in Computational Intelligence*, *Advances in Intelligent Systems and Computing*, Springer, Singapore, vol. 1176.

Sitaula, N., Mohan, C.K., Grygiel, J., Zhou, X., & Zafarani, R. (2020, June 18). Credibility-based fake news detection. In K. Shu, S. Wang, D. Lee, & H. Liu (Eds), Disinformation, misinformation, and fake news in social media, *Lecture Notes in Social Networks*. Springer, Cham, 163-182.

Stavrianou, A., Andritsos, P., & Nicoloyannis, N. (2007, September). Overview and semantic issues of text mining. *SIGMOD Record*, 36(3), 23-34.

Visser, J., Lawrence, J., & Reed, C. (2020, November). Reason-checking fake news. *Communications of the ACM*, 63(11), 38-40.

Zhou, X. & Zafarani, R. (2020, October). A survey of fake news: fundamental theories, detection methods, and opportunities. *ACM Computing Surveys*, 53(5):109, 1-40.